# Current measurement by real-time counting of single electrons

Jonas Bylander, Tim Duty and Per Delsing

*Department of Microtechnology and Nanoscience (MC2), Chalmers University of Technology, SE-412 96 Göteborg, Sweden*

**The fact that electrical current is carried by individual charges has been known for over 100 years, yet this discreteness has not been directly observed so far. Almost all current measurements involve measuring the voltage drop across a resistor, using Ohm's law, where the discrete nature of charge does not come into play. However, by sending a direct current through a microelectronic circuit with a chain of islands connected by small tunnel junctions, the individual electrons can be observed one by one. The quantum mechanical tunnelling of single charges in this one-dimensional array is time correlated[1-3], and consequently the detected signal has the average frequency $f=I/e$, where $I$ is the current and $e$ is the electron charge. Here we report a direct observation of these time correlated single-electron tunnelling oscillations and show electron counting in the range 5 fA–1 pA. This represents a fundamentally new way to measure extremely small currents, without offset or drift. Moreover, our current measurement, which is based on electron counting, is self-calibrated, as the measured frequency is related to the current only by a natural constant.**

In the mid 80's, it was suggested[1] that a small current consisting of individual electrons, tunnelling through a small tunnel junction, could at low temperatures result in an oscillating voltage of amplitude $e/C$, where $C$ is the capacitance of the tunnel junction. The full theory for these so-called single-electron tunnelling oscillations was then developed[2], based on earlier work on Bloch oscillations and the underlying Coulomb blockade[4,5]. This phenomenon of single-electron tunnelling oscillations is similar to the a.c. Josephson effect, as phase and charge are quantum conjugated variables. However, the duality is not complete because the single-electron tunnelling is lacking phase coherence. A few years later these oscillations were detected indirectly by phase locking to an external microwave signal[6]. Shortly thereafter, new devices such as the single electron turnstile[7] and the single electron pump[8] were invented in order to create a current given by the fundamental relation $I=ef$. Since then, the single electron pump has been refined to a very high accuracy[9].

A number of authors have also proposed[10-13] that it should be possible to turn this relation around and instead measure the current by monitoring the individual electrons as they pass through a circuit. More recently, single electron tunnelling events have been observed[14,15]. In those experiments, however, there was no time correlation, and thus no relation between frequency and current could be demonstrated.

In order to measure current by electron counting, three main ingredients are necessary: (i) time correlation of the tunnelling events, (ii) a fast and sensitive charge detector, and (iii) a very stable current bias. To bring about time correlation in a single tunnel junction, in contrast to uncorrelated shot noise[16], care must be taken to make the electromagnetic impedance seen by the junction large[2] compared to the Klitzing resistance $R_K = h/e^2 = 25.8$ kohm. This can be achieved by placing small-size resistors in close proximity to the junction[17,18] or by using a one-dimensional series array of tunnel junctions[19].

In this experiment, we have used a superconducting array containing $N=50$ junctions (Figure 1). The capacitance of each junction is $C_A = 0.42$ fF, and the stray capacitance of an electrode inside the array is $C_0 = 30$ aF. In such an array excess charge on one island polarizes the neighbouring islands, so that the charges repel each other. The charge is localized, but the potential resulting from the capacitive voltage division in the array is spread over a number of junctions $M$. This potential distribution is often referred to as a charge soliton[3] (or antisoliton, if a charge is missing), since it can move (by tunnelling) throughout the array without changing its form. In our case the soliton length is $M=(C_A/C_0)^{1/2} = 3.7$, and the array is in the long limit, $N>>M$. When the array is biased above a certain threshold voltage, $V_t$, charges enter from the edge, and as the solitons in the array repel one another, a moving quasi-Wigner lattice of charge solitons is formed. It was shown theoretically[3,20], and observed indirectly[6], that this type of array exhibits time correlated tunnelling events.

In order to detect the tunnelling charges in real time, we have used an improved version of the conventional single-electron transistor[2,21,22] (SET) as a charge sensor, namely the radio-frequency SET[23] (Fig. 1). This is the only device with the required sensitivity and speed: it can detect sub-electron charge changes at frequencies well above 10 MHz. One possibility to do electron counting, as suggested by Visscher[13], is to couple the SET capacitively to one of the electrodes inside the array, and thus observe the potential variations of that electrode as the charge solitons pass. However, we have instead chosen to couple the array directly to the middle electrode (island) of the SET as can be seen in Figure 1, a configuration known as the resistively coupled[2,24,25] or array-coupled[26] SET. In this way, the full electron charge is detected by the SET.

We have designed the device parameters in order to obtain the desired soliton behaviour and at the same time ensure good performance of the SET. The characteristic time for the soliton dynamics is determined by the $RC$ constant of the tunnel junctions in the array. Monte Carlo simulations[3,27] show that the narrowest linewidth of the single-electron tunnelling oscillations occurs at approximately one percent of the characteristic $e/RC$ current. To keep this current low enough for the SET to follow the oscillations, it is important that the array tunnel junctions have a high resistance. For the SET, on the other hand, sensitivity falls off as the resistance increases. To obtain different oxide thickness for the array and SET junctions we have therefore employed a three-angle shadow evaporation process to deposit aluminium[28]. For the sample described here, the SET source-drain resistance was $R_{SET} = 30$ kohm, while the array resistance was $R_N = 0.94$ Mohm per junction. In Figure 2, we show the array current-voltage characteristics.

The measurements presented here are done at a parallel magnetic field $B_\parallel = 475$ mT, where the array is still superconducting. This makes the system less sensitive to bias voltage fluctuations, since in the superconducting state the onset of current is rounded, while in the normal state it is quite steep (Figure 2 inset). Moreover, the sensitivity of the SET is higher in the superconducting state. In general, both electron and Cooper pair tunnelling can take place in the array, but at this magnetic field the threshold for electron injection is lower than that for Cooper pairs. The suppressed gap, $Delta/k_B = 0.6$ K, where $k_B$ is Boltzmann's constant, and the high impedance of the array junctions also suppress the tunnelling rates for Cooper pairs, and we expect to see only electrons at this field.

When the array is biased slightly above the threshold voltage for electrons, $V_t^e$, a small (sub-pA) current starts to flow. As the charges approach the SET, the source-drain current is modulated by the charge induced on the middle island, resulting in modulation of the reflected RF power. In this way, we have detected single-electron tunnelling oscillations in real time. Figure 3a shows time traces of single-electron tunnelling events for three different currents. In the range from 5 fA to 1 pA, the reflected power spectrum shows a peak (Figure 3b) at a frequency that corresponds to the applied current in the array, $f=I/e$ (Figure 4). This demonstrates time correlation between consecutive tunnelling events. The signal is gradually smeared by temperature, but persists even at 300 mK. The peak broadens considerably when the current approaches 10 % of the $e/RC$ current, in fair agreement with simulations, assuming a subgap resistance $R_{SG}=100R_N=94$ Mohm.

Our preamplifier, with a noise temperature of 2 K, sets the overall noise level in our measurement system. The counting errors caused by the limited sensitivity of the SET can be estimated[13] as $P=0.5\text{erfc}(q_c/Q_N2^{1/2})$, where $Q_N$ is the charge noise within the measurement bandwidth, $q_c$ is the charge threshold for a count to be registered, and erfc is the complementary error function. For realistic parameters, this expression gives an accuracy of 1 part in $10^6$. However, $P$ is very strongly dependent on the sensitivity of the SET, which suggests that metrological accuracy can be achieved in an optimized device. In this experiment, there are obviously other error sources limiting the accuracy, such as bias instability that smears the peak in the spectrum and background charges that affect both the operation point of the SET and the soliton flow in the array.

An important property of the electron counter described in this letter is its dynamic current range. At low currents (<3 fA, according to our simulations[27]), when there is only one soliton in the array at a time, space correlation, and hence also time correlation, is lost. The opposite limit is ultimately set by the $e/RC$ current. In a realistic experiment, however, the resolution bandwidth of our measurement set-up, $1/f$ noise and instability of the bias at very low currents prevent us from measuring with any accuracy below 5 fA. At high currents, the measurement speed is restricted by the bandwidth of the SET.

From an applications point of view it would be desirable to measure larger currents, and to decrease the input impedance of the ammeter. This is possible with an electron counter, since it is straightforward to put several devices in parallel, which is not the case with single-electron pumps. Then it can also be used to close the quantum metrological triangle[5] relating current, voltage and frequency by fundamental constants. Another important feature is that the counter is self calibrated, and does not suffer from problems like offset or drift in the measurement set-up.

Recent interest in quantum noise in mesoscopic systems has lead to the theory of full counting statistics[29], in which general information about charge transfer can be gained from the higher statistical moments of noise. So far, very few experiments have been performed along these lines, but with the relatively low bandwidth of the tunnelling process in the array we see a possibility to measure the full statistics of the charge transport and get access to higher statistical moments of the current.

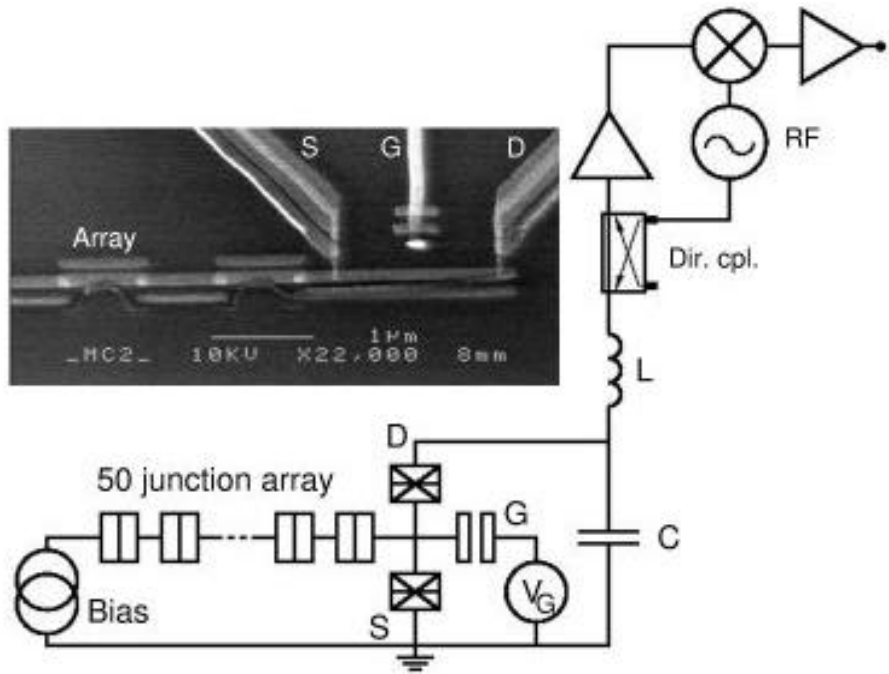

Figure 1. Experimental set-up. Scanning electron micrograph of the sample and a schematic layout of the electron counter based on an SET. Via a directional coupler, an RF signal ($f$=358 MHz) is applied to the $LC$ circuit in which the SET is embedded. The quality factor of the resonator is $Q$=15, giving a bandwidth of about 10 MHz. The reflected power is sensitively dependent on the charge on the SET island (the metallic strip connected to the source (S) and drain (D) electrodes of the SET and to the array). For tuning the working point of the SET, a voltage is applied to the capacitive gate (G). The reflected signal is first preamplified, and then mixed with a local oscillator. We measured the small-signal charge sensitivity of the SET to be $2 \cdot 10^{-5}$ $e$Hz$^{-1/2}$ by applying a 200 kHz sine wave of amplitude 0.02 $e_{\mathrm{rms}}$ to the gate G (at $B_{\parallel}$=475 mT). The charging energy of the SET is $E_C/k_B=e^2/2C_{\mathrm{sum}}k_B$=1.6 K, where $C_{\mathrm{sum}}$ is the total capacitance of the island. Either voltage bias (using a transimpedance amplifier Stanford Research 570 with a dc source) or current bias (with a Keithley 263 Calibrator/Source) can be applied to the array of 50 very small tunnel junctions in order to create a current. Injecting the charge directly into the SET island ensures very good coupling to the measurement device. The measurements were performed in a dilution refrigerator at $T$=30 mK.

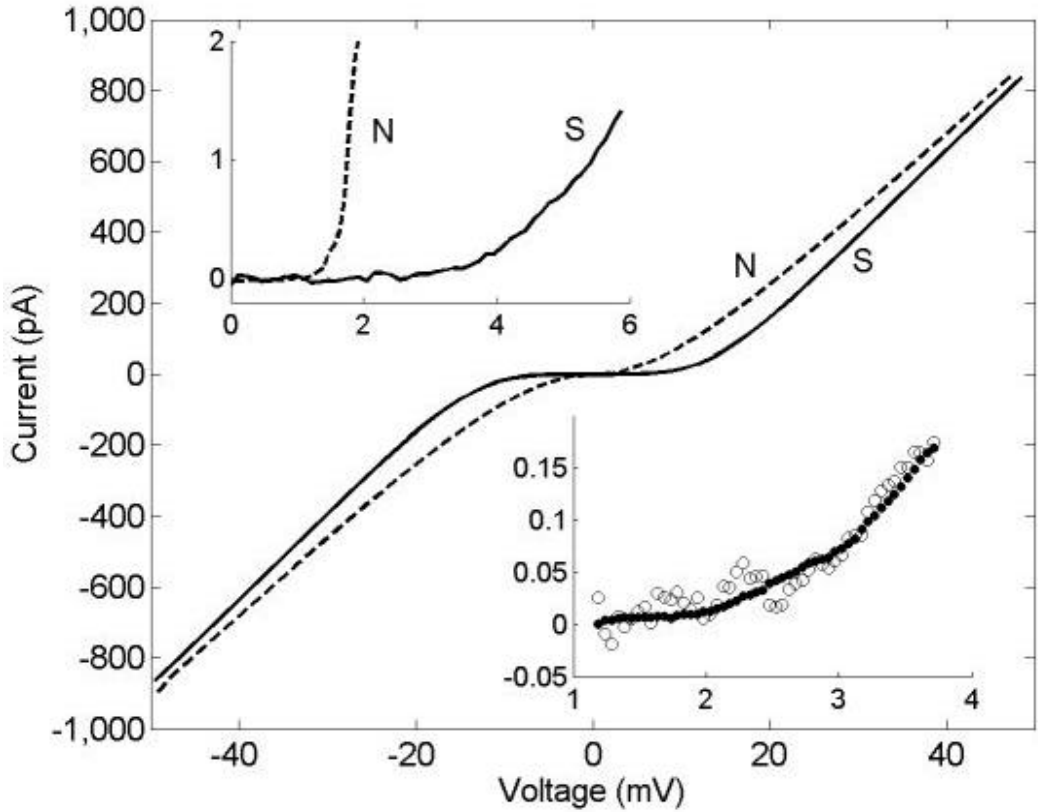

Figure 2. Array current-voltage characteristics in the superconducting state ($B_\parallel$=475 mT, solid line) and in the normal state (1.50 T, dashed line). The critical field is $B_{\parallel,C}$=650 mT. The asymptotic normal resistance is 47 Mohm, and the offset voltage $V_{off}$=190 μV per junction gives the charging energy $E_C^{Array}/k_B$=e$^2$/2$C_A k_B$=2.2 K.

Insets have the same units as the main graph. Upper inset: Above threshold, the current onset is sharp in the normal state, but more gradual in the superconducting state due to the subgap resistance $R_{SG}$, which is both magnetic field and voltage dependent and much higher than the normal resistance $R_N$. Consequently, the sensitivity to bias voltage fluctuations is much higher in the normal state. Ignoring the effect of background charges, the threshold voltage for electrons is theoretically $V_t^e$=$KMe$/2$C_A$+(1+$M$)$Delta(B)/e$=0.87 mV, where $K$=((1+1/4$M^2$)$^{1/2}$+1/2$M$)$^{-1}$=0.9, whereas for Cooper pairs it is $V_t^{2e}$=$KMe$/$C_A$=1.2 mV. The effect of random background charges further complicates the situation by altering these thresholds[30,31]. Due to the weak Josephson coupling energy, $E_J/k_B$=1 mK<<$T$, and the high impedance of the neighbouring junctions (we assume a subgap resistance $R_{SG}$=100$R_N$), the rate for Cooper pair tunnelling is much lower than the rate for electron tunnelling also above $V_t^{2e}$.

Lower inset: Comparison of d.c. current measurement by counting individual electrons, *i.e.*, measuring a frequency, (solid dots) and a conventional d.c. current measurement (open circles) by voltage biasing through a Stanford

Research 570 transimpedance amplifier. Note that for very low currents, the spread in the frequency measurements is smaller than in the conventional current measurements.

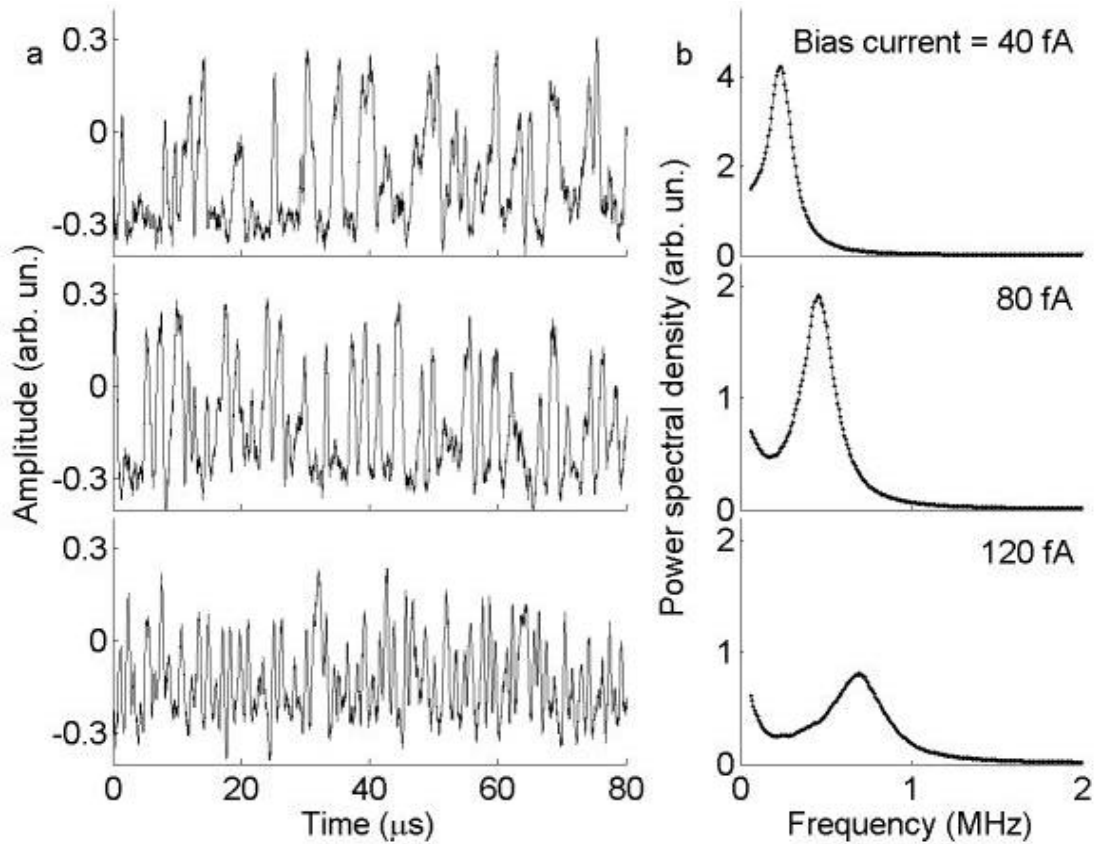

Figure 3. Experimental data. **a**, Real time electron counting. Each peak in the reflected signal corresponds to one electron tunnelling into the SET island. The traces are snapshots recorded on a digital oscilloscope. The signal is low-pass filtered with a cut-off frequency of 15 MHz and then digitally smoothed with a moving average of 0.5 µs. The applied currents were, from top to bottom, 40, 80 and 120 fA, corresponding to the frequency (*f=I/e*) 250, 500 and 750 kHz, respectively.

**b**, Power spectral densities of the reflected power from the SET, measured using a spectrum analyzer with 100 kHz resolution bandwidth. The peak frequencies are *f*=236, 460 and 700 kHz, corresponding to the currents 38, 74 and 112 fA, respectively. The discrepancy from the nominal bias currents is due to a small offset in the current source.

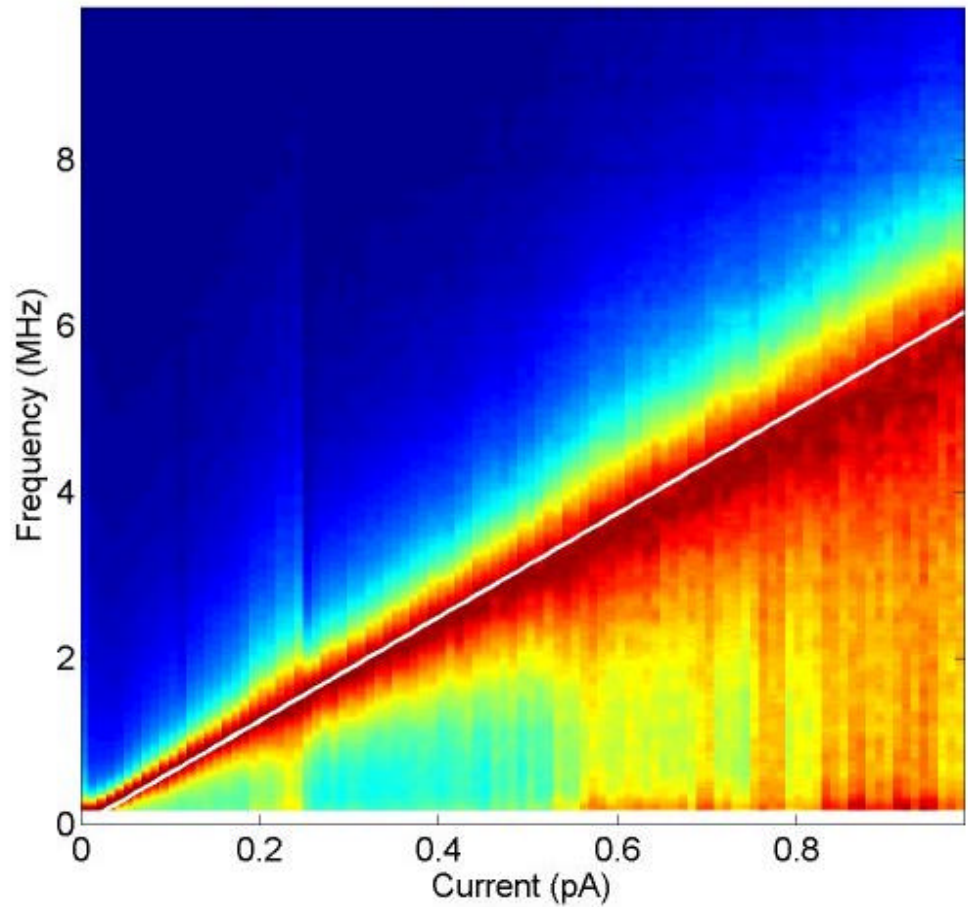

Figure 4. Single-electron tunnelling oscillations. The power spectra (Fig. 3b) have been normalized so that the peaks all have unity height. The dynamic colour range goes from blue to red, corresponding to zero and one, respectively. The position of the red ridge agrees well with the solid white line $f=I/e$, where $I$ is the applied bias current. This demonstrates current measurement by electron counting from 5 fA to 1 pA. The noticeable broadening and increasing asymmetry of the spectral peak for higher currents gives a small discrepancy between the measured data and $I/e$.

**Acknowledgements** We would like to thank C. Kristoffersson, S. Pedersen and P. Wahlgren for assistance in the early stages of this work, K. Bladh, D. Gunnarsson, S. Kafanov and M. Taslakov for technical assistance, and T. Claeson, H. Nilsson, K-E. Rydler, G. Wendin, C. Wilson and A. Zorin for discussions. The work was supported by the Swedish SSF and VR, the EU research project COUNT and by the Wallenberg foundation.

**Correspondence** and requests for materials should be addressed to J.B. (jonas.bylander@mc2.chalmers.se).